\pdfoutput=1

\documentclass[11pt]{article}

\usepackage{acl}

\usepackage{times}
\usepackage{latexsym}

\usepackage[T1]{fontenc}

\usepackage[utf8]{inputenc}

\usepackage{microtype}

\usepackage{amsmath}
\usepackage{amsfonts}

\usepackage{inconsolata}
\usepackage{multirow}
\usepackage{graphicx}
\usepackage{subcaption}

\title{A Robust Semantics-based Watermark for Large Language Models \\ against Paraphrasing}

\author{Jie Ren$^{1}$, Han Xu$^{1}$, Yiding Liu$^{2}$, Yingqian Cui$^{1}$, Shuaiqiang Wang$^{2}$ \\ \textbf{Dawei Yin$^{2}$,} \textbf{Jiliang Tang$^{1}$} \\
$^{1}$Michigan State University, $^{2}$Baidu Inc. \\
\texttt{\{renjie3, xuhan1, cuiyingq, tangjili\}@msu.edu} \\
\texttt{liuyiding.tanh@gmail.com, shqiang.wang@gmail.com, yindawei@acm.org}
  }

\begin{document}
\maketitle
\begin{abstract}
Large language models (LLMs) have show their remarkable ability in various natural language tasks. However, there are concerns that LLMs are possible to be used improperly or even illegally. To prevent the malicious usage of LLMs, detecting LLM-generated text becomes crucial in the deployment of LLM applications. Watermarking is an effective strategy to detect the LLM-generated content by encoding a pre-defined secret watermark to facilitate the detection process. However, the majority of existing watermark methods leverage the simple hashes of precedent tokens to partition vocabulary. Such watermarks can be easily eliminated by paraphrase and, correspondingly, the detection effectiveness will be greatly compromised. Thus, to enhance the robustness against paraphrase, we propose a semantics-based watermark framework, SemaMark. It leverages the semantics as an alternative to simple hashes of tokens since the semantic meaning of the sentences will be likely preserved under paraphrase and the watermark can remain robust. Comprehensive experiments are conducted to demonstrate the effectiveness and robustness of SemaMark under different paraphrases. Our code is available at \href{https://github.com/renjie3/SemaMark}{github.com/renjie3/SemaMark}. 
\end{abstract}

\section{Introduction}

Large language models (LLMs) have shown their great ability in various natural language processing (NLP) tasks like Question Answering (QA)~\cite{lu2022learn}, reasoning tasks~\cite{wei2022chain, creswell2022selection} and code development~\cite{xu2022systematic}. However, tremendous concerns have been raised that LLMs are possible to be used improperly and illegally. For example, indistinguishable fake news are easy to be fabricated~\citep{kreps2022all, zellers2019defending} by language models, which, when disseminated, could instigate widespread panic. 
Similarly, in the commercial sphere, convincingly generated reviews can manipulate consumer perceptions, leading to unethical business competition~\citep{salminen2022creating}. 
Therefore, detecting LLM-generated text has become crucial in the real-world applications of LLMs~\citep{wu2023survey, sadasivan2023can, xu2023generalization}.

{Among diverse methods to detect LLM-generated texts}, the watermark strategies have demonstrated outstanding precision~\citep{liu2023survey, tang2023science, ren2024copyright}. It is proposed to encode a secret watermark into the generated texts, such that we can tell whether a text is generated by detecting this watermark. One representative strategy~\cite{kirchenbauer2023watermark, yoo2023robust} is to encode the watermark based on the ``partition of vocabulary''. In detail, given a language model, these methods devise a mapping from precedent tokens to a particular partition of the vocabulary by a partition function for the consequent token. The partition function leverages the hashes of the input as the seed of a random generator to split the vocabulary to a green list and a red list. During the text generation phase, the consequent token has an increased probability to be sampled from the green list. In this way, the watermark is encoded through the matching between the precedent tokens and the vocabulary partition for the consequent token. The detection is also facilitated by detecting this matching in generated contents. However, recent works~\cite{krishna2023paraphrasing, kaddour2023challenges} reveal that \textbf{this watermark may be easily eliminated by sentence paraphrasing}. 
Individuals seeking to {improperly} utilize LLMs without being detected can paraphrase the generated contents, like altering the order and the choices of the words, and only retain the general meaning of the text {to achieve their malicious goals like faking news}. 
These paraphrases will change the seed of the partition function, i.e. the token hashes, and as we show in the Section~\ref{sec:alabtion}, the partition function is sensitive to small changes. Consequently, the matching between the precedent tokens and the green list will be disrupted, and the detection effectiveness of the watermark can be dramatically compromised.

In this paper, we propose to leverage the semantic meaning of precedent token sequences as the seed for partition function, instead of simple hashes of precedent tokens, since the core semantic meaning is expected to be maintained after paraphrase. To achieve this goal, one key obstacle is how to capture the semantics when applying them for the partition function to watermark the generated texts. It is a common practice to quantify the semantics via embeddings~\cite{reimers2019sentence, gao2021simcse, li2020sentence, giorgi2021declutr}. Embeddings indeed can represent consistent semantics after paraphrase. Since the embeddings are high-dimensional vectors in the continuous space, they often present some minor changes after paraphrase. 
Although the main semantics are preserved, these minor changes can lead to a substantial difference in the partition of vocabulary because the random generator in the partition function is sensitive to the change of the seed, as shown in Section~\ref{sec:alabtion}.


To overcome the above challenge, i.e., to make the quantified semantics invariant and make the watermark robust under paraphrase, we propose a new watermark method, SemaMark, which discretizes the continuous embedding space. Intuitively, the discretization can coarsen the representation of the embeddings which could tolerate the potential minor changes caused by paraphrase. By proper discretization, the paraphrased semantics could stay in the same discrete section with a high probability and the discretized quantified semantics will likely remain the same even after paraphrase. Therefore, the partition results will not change. However, directly converting the high-dimensional embedding space into discrete is intricate and challenging. For example, discretizing each dimension will lead to a large amount of discrete values which is exponential to the number of dimensions.
Thus, the minor changes by paraphrase can still cause the change of discrete values because the number of discrete values are too dense and each discrete value can tolerate only small changes.
Therefore, the minor changes of high-dimensional embeddings can have a strong impact on the partition function. 
To address this problem, SemaMark first uses a Multi-Layer Perception (MLP) to condense the continuous high-dimensional embeddings into normalized vectors in 2D space. The vectors are located on a unit circle named Normalized Embedding Ring (NE-Ring). Then the condensed NE-Ring is equally divided into various sections, transforming the continuous space into distinct discrete values, i.e., ``semantic values''. Based on the discretization, SemaMark further introduces two strategies to advance the watermark's concealment and to improve the robustness under paraphrase. 
\textit{First}, SemaMark leverages the uniformity~\cite{wang2020understanding} of Contrastive Learning(CL)~\cite{chen2020simple} to strength the MLP and mitigate the problem that the semantics are unevenly concentrating on some discrete sections on NE-Ring. The unevenly distribution will cause the final discrete semantic values overly monotonous. It raises the concern that the watermark might be cracked by counting token frequency~\cite{zhao2023provable}. \textit{Second}, SemaMark utilizes an offset detection method to further enhance the robustness at the boundary of different discrete sections whose semantic values are possibly vulnerable to paraphrase. Comprehensive experiments are conducted to demonstrate the effectiveness and robustness of SemaMark under different paraphrases. 




\section{Related works}

\paragraph{LLM-generated detection.}
As the development of LLMs, various LLM-generated detection tools have also been proposed. Learning-based methods train a classification model to detect the difference between human-written text and machine-generated text like ~\citet{guo2023close, wang2023seqxgpt, li2023origin}. Other works do not rely on the classification model, but try to use the property of the LLM to test whether a given text is generated by LLMs. For example, DetectGPT~\cite{mitchell2023detectgpt} assumes that the generated text will have high likelihood. GPT-who~\cite{venkatraman2023gptwho} uses UID-based features to model the unique statistical signature of each LLM and human author for accurate authorship attribution. These methods do not interact the generation process of LLMs and thus have to explore unknown features of LLMs for detection. Instead, watermarks can change the model with a small but pre-defined rule which accelerates the detection process effectively.

\begin{figure*}[th]
  \centering
  \includegraphics[width=1\textwidth]{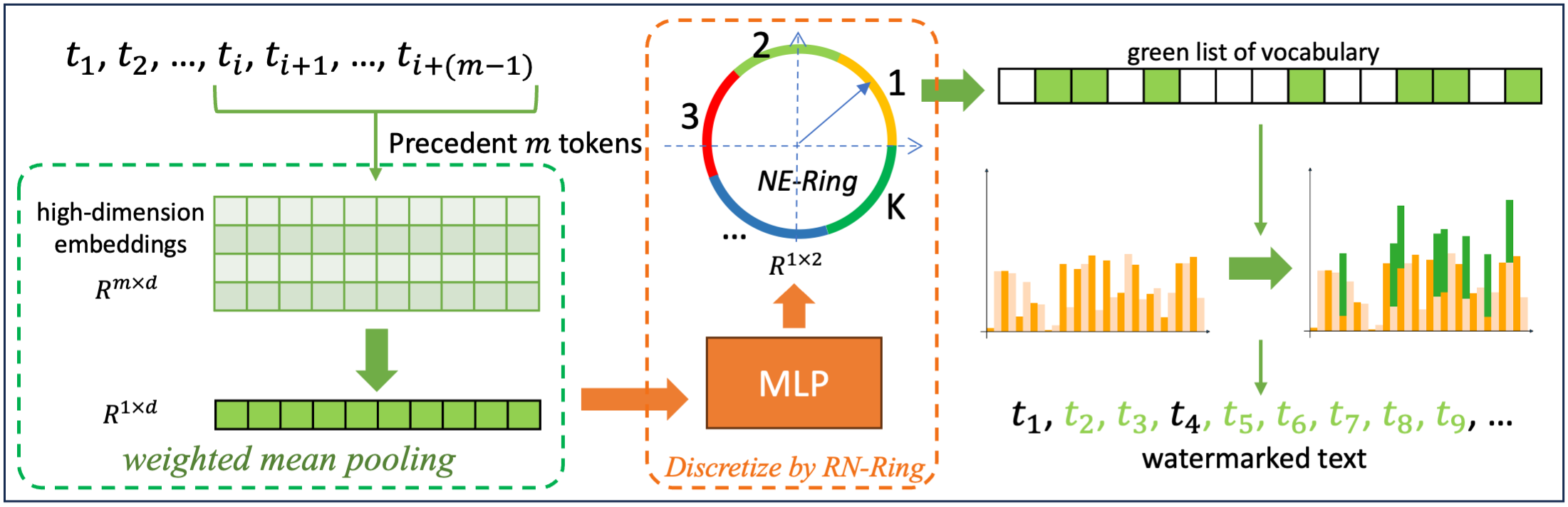}
  \caption{The watermarking process of SemaMark}
  \label{fig:framework} 
\end{figure*}

\paragraph{Watermark.} The distinction between watermark and other methods is that watermark can proactively change the generation to insert a concealed watermark into the generated text. This gives clear difference between watermerked and non-watermarked texts.
Watermark shifts the text using a small but pre-defined rule to make the detection much more effective. The partition of the vocabulary for each token is a representative watermark method~\cite{kirchenbauer2023watermark, yoo2023robust, kirchenbauer2023reliability}. In each auto-regressive step of generating one token, the method uses the previous tokens' hashes, to select a part of the vocabulary as ``green'' at a ratio of $\gamma$. Subsequently, they elevate the likelihood of the tokens by boosting the logits of the softmax by $\delta$. Through this approach, at each token position, the probability of this matching between the seed and green tokens tends to increase. 

For a sentence with $L$ tokens, it is viewed as a sample set of size $L$. Each token is one sample from the vocabulary. A non-watermarked sentence is expected to have $\gamma L$ tokens showing this match, while the watermarked sentence is expected to have more. The watermark detection is approached as a $z$-test with null hypothesis that the text is non-watermarke. If the $z$-statistic is large, i.e. it is significantly different from the null hypothesis, the null hypothesis can be rejected and the text can be predicted as watermarked:
\begin{align}
    z=\frac{\left(G-\gamma L\right)}{\sqrt{L \gamma(1-\gamma)}},
\label{eq:z_score}
\end{align}
where $G$ is the number of tokens showing the matching between seed and the green list. \citet{yoo2023robust} further expand this watermark of green and red list to more lists for multi-bit encoding.

\cite{liu2023semantic} propose a semantic invariant method to watermark the generated text of LLM. However, their method employs two additional models, introducing redundant encoding processes in the text encoder, which can be time-consuming.

\section{Method}
\label{sec:method}

In this section, we introduce the detailed design of SemaMark. We first present how to use the semantic information as the seed for watermark methods that are based on random partition of vocabulary in Section~\ref{sec:main_method}. Then in Section~\ref{sec:cl} and Section~\ref{sec:effset_detection}, we introduce the CL training scheme and the smoothed detection method for further improving the robustness, respectively.

\subsection{The framework of SemaMark}
\label{sec:main_method}

{As aforementioned, the existing watermark methods based on partition of vocabulary are susceptible to paraphrase. Paraphrase can easily change the previous tokens and disrupt the matching between tokens and the partition of vocabulary, without significantly affecting the semantic meaning.} Thus, SemaMark uses the invariant semantics for watermarking by discretizing the embedding space to accommodate the minor perturbation of semantics and provide a stable mapping between semantics and vocabulary partition for the consequent token.

However, discretization in a high-dimension space is intricate and non-trivial. Therefore, we first reduce the high-dimensional embedding space onto the 2D NE-Ring and then discretize via NE-Ring. The whole watermarking process is shown in Figure~\ref{fig:framework}. SemaMark first reduces the dimension of the embedding space to obtain the discrete semantic values by two steps, i.e., \textit{weighted embedding pooling} and \textit{discretizing by NE-Ring}, and then uses the semantic value to partition the vocabulary. The logits of green list is shifted to increase the probability of matching between semantics and the consequent token for watermarking the LLM, $f$.
In the following, we introduce more details about the two steps to obtain a stable semantic value.


\paragraph{S1: weighted embedding pooling.} 
To enhance the robustness, we aggregate the semantics of previous $m$ tokens by the weighted mean pooling function $P(\cdot)$ before dimension reduction, instead of using only one preceding token's embedding. In the ablation studies of Section~\ref{sec:alabtion}, we show that the method has the best performance when $m$ is neither too big nor too small. For the token sequence $\{t_{i:i+m-1}\}$ starting at position $i$, we use their semantics to generate the token in the $m$ position, $t_{i+m}$. We denote their embeddings as $\{\boldsymbol{e}_{i:i+m-1}\}$. $\{\boldsymbol{e}_{i:i+m-1}\}$ can be easily obtained from the LLM, $f$, that we want to watermark. 
Intuitively, in $\{t_{i:i+m-1}\}$, the embeddings of tokens far from $t_{i+m}$ contain semantic information that is more distant from $t_{i+m}$ than the closer ones.
The connection between distant tokens might be more possible to change after paraphrase compared with closer tokens. Thus, in the sequence $\{t_{i:i+m-1}\}$, the embeddings of distant tokens might be less robust. To increase the robustness for the green list of the current token position $t_{i+m}$ after paraphrase, the pooling embeddings should rely more on the closer tokens, therefore, we use a linear weight function to assign lower weights to tokens far from $t_{i+m}$ and higher weights to those in closer proximity:
\begin{align*}
    P(\{\boldsymbol{e}_{i+1:i+m}\}) = \sum_{j=1}^{K} \frac{j + \frac{K}{2}}{w_{\text{sum}}} \boldsymbol{e}_{i+j},
\end{align*}
where $w_{\text{sum}} = K^2 + K/2$ is the sum of all weights. We denote the weighted output $P(\{\boldsymbol{e}_{i:i+m-1}\}) \in \mathbb{R}^{d}$ as $\boldsymbol{e}_{P_{i,m}}$ for short. By pooling, more semantics are used for a seed, which enhance the robustness under paraphrase. 

\paragraph{S2: discretizing by NE-Ring.} After aggregating the embeddings by weighted pooling, SemaMark uses MLP $g_{\theta}$ to transform $\boldsymbol{e}_{P_{i,m}}$ to a normalized vector in 2D embedding space. The normalized vectors locate on a unit circle in the 2D space, which is named as Normalized Embedding Ring (NE-Ring). 
The discretization function, $D(\cdot)$, discretizes NE-Ring by equally segmenting into different sections. It takes the polar angle $\phi$ of $g_{\theta}(\boldsymbol{e}_{P_{i,m}})$ as input and outputs the discretized semantic values $a \in [K]$, where $[K]:= \{1, 2, ..., K \}$. $D(\cdot)$ is defined as
\begin{align*}
    D(\phi)=\left\lfloor\phi\frac{K}{2 \pi}\right\rfloor
\end{align*}
It first maps the input from $[0, 2\pi)$ to $[0, K)$, and then discretizes all the values in $[i, i+1)$ to $i$, for $\forall i \in [K-1]$. Even though there could be subtle changes in semantics by paraphrase, the paraphrased $\tilde{a}$ will likely locate in the discrete section $[i, i+1)$. Some tokens may still have $a \neq \tilde{a}$ if the normalized vector is close to the boundary of $[i, i+1)$. Therefore, in Section~\ref{sec:effset_detection}, we introduce an offset detection to strengthen the tolerance for this mismatch and correct some unstable cases.

With the two steps, we can get a stable discrete semantic value as the seed for the partition function to partition the vocabulary for the consequent token. Following ~\citet{kirchenbauer2023watermark}, the vocabulary is partitioned into green and red lists. We increase the logits of the tokens in the green list by $\delta$ and recalculate the probability distribution based on the shifted logits. For each token to generate, we increase the possibility of the green list based on its previous $m$ tokens' semantics. Thus, all the generated tokens will be likely to have this matching between the semantics and the consequent green token. By detecting the matching, we can discriminate whether a text is watermarked or not and then detect the LLM-generated contents effectively. Besides, SemaMark proposes two strategies to reduce the risk of being cracked by Contrastive Learning and further increase the robustness by the offset detection in the following sections.



\subsection{Training $g_{\theta}$ by Contrastive Learning}
\label{sec:cl}
%
The MLP is expected to produce a uniform distribution of $g_{\theta}(\boldsymbol{e}_{P_{i,m}})$ on NE-Ring. If different semantics unevenly distributed on NE-Ring, the resulting discrete semantic values will be overly monotonous and the green list is more changeless. Consequently, the green list might be revealed by counting the token frequency, which compromises the concealment of watermark and leads to the risk of being cracked. Ideally, SemaMark should generate a wider variety of semantic values for different sentences, while each semantic value is robust and stable if its corresponding sentence is paraphrased.
To achieve this goal, we propose to use Contrastive Learning to train MLP since 
Contrastive Learning has the property of uniformity that the data will be evenly distributed in the whole feature space~\cite{wang2020understanding}. 
The uniform distribution can help the normalized vectors cover all the semantic values. As a result, NE-Ring can generate a wider variety of semantic values to prevent the watermark from being cracked.


In Contrastive Learning, we first input the sentences into the model $f$ to get a batch of sequences of $m$ tokens and their pooling embeddings $\boldsymbol{e}_{P_{i, m}}$, denoted as $\{\boldsymbol{e}_{j}\}$, where $j \in [B]$ and $B$ is the batch size.
To compose a contrastive loss, we construct the positive and negative pairs by a soft augmentation:
\begin{align*}
    \boldsymbol{e}_{j+B} = \boldsymbol{e}_{j}^{+} = \boldsymbol{e}_{j} + \epsilon,
\end{align*}
where $\epsilon \sim \mathcal{N}(0,\sigma^2)$ is a Gaussian noise. The soft augmentation can simplify the construction of positive samples. With this soft augmentation, we can assign the samples sharing similar embeddings from the same sequence as positive pairs and samples from different sequences as negative pairs. This is consistent with our intuition that the paraphrased semantic embeddings will not change significantly and can remain robust.
Then the contrastive loss is 
\begin{align*}\small
    L_{j}=-\log \frac{\exp \left(\operatorname{sim}\left(g_{\theta}(\boldsymbol{e}_j), g_{\theta}(\boldsymbol{e}_j^{+})\right) / \tau\right)}{\sum_{k\neq j, k \in [2B]} \exp \left(\operatorname{sim}\left(g_{\theta}(\boldsymbol{e}_j), g_{\theta}(\boldsymbol{e}_k)\right) / \tau\right)},
\end{align*}
where $\operatorname{sim}(\cdot)$ is cosine similarity and $\tau$ is the temperature. By Contrastive Learning, the output of reduced semantic embeddings can be evenly distributed in all of the space on NE-Ring, and cover all the discrete sections to improve the robustness of SemaMark.



\subsection{$Q$-offset detection}
\label{sec:effset_detection}


\begin{figure}[th]
  \centering
  \includegraphics[width=0.45\textwidth]{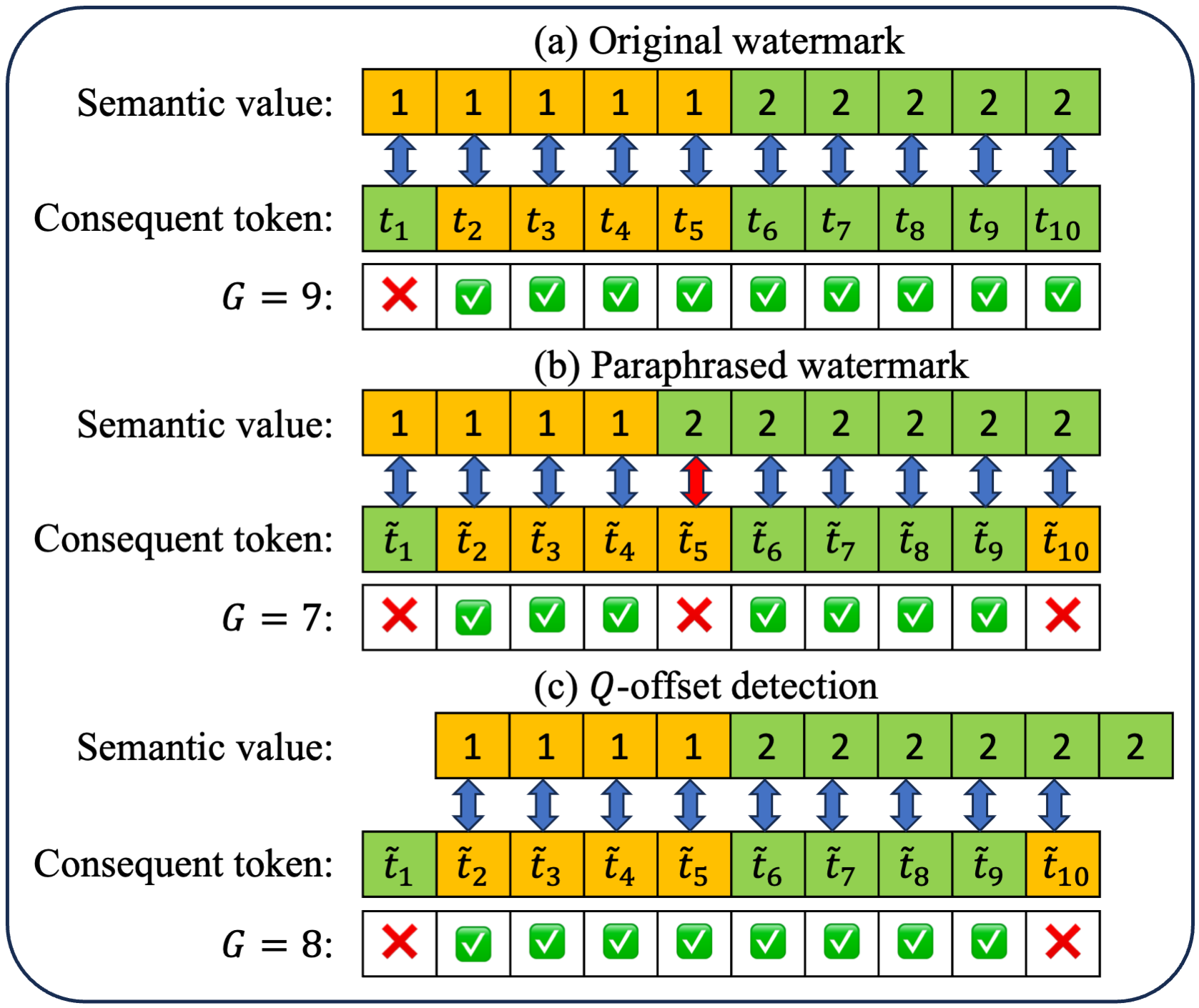}
  \caption{$Q$-offset detection vs. existing detection}
  \label{fig:offset_ex} 
\end{figure}

\noindent Existing detection methods check the matching between partition seed and the consequent tokens in a one-to-one manner as shown in Figure~\ref{fig:offset_ex}(a). The detection method first recalculates the seed for each token position and gets the partition of the green list, and then checks whether the consequent token is in the partitioned green list token by token. In SemaMark, this strategy can be effective when the text is not paraphrased. However, after paraphrase, this detection could be suboptimal because the semantic values of some sequences which are close to the boundaries of the discrete section $[i, i+1)$ might change as shown in Figure~\ref{fig:offset_ex}(b). This is because the window of $m$ tokens will slide token by token during the auto-regressive generation process, and the semantic change will also accumulate when the window is sliding. The semantic values closed to the boundary usually happen when the change accumulates to some extent. This change of boundary semantic values will lead to some mismatch and reduce the accuracy like $\tilde{t}_5$ in Figure~\ref{fig:offset_ex}(b).



To mitigate the influence of this error, we propose $Q$-offset detection. As shown in Figure~\ref{fig:offset_ex}(c), we offset the discrete seed by $q$ tokens to detect the matching between semantics and the consequent tokens, where $q \in \{-Q, -(Q-1), ..., 0, 1, ..., Q\}$ and the sign of $q$ indicates the direction of the offset. We choose the maximal $z$-statistic in different $q$ as the $Q$-offset score. However, $Q$-offset detection will also increase the $Q$-offset score of non-watermark text, which indicates that the detected green word fraction $\gamma$ of non-watermark text is higher. The $\gamma$ in Eq.~\eqref{eq:z_score} is possibly inaccurate. Thus during generation, we set $\gamma$ to a fixed value, while in detection process, we treat $\gamma$ as a hyper-parameter and use a validation set to determine its value in practice. In Section~\ref{sec:alabtion}, we discuss the ablation studies of $Q$-offset and $\gamma$ and show that $Q$-offset can impressively improve the detection performance with robustness.

\section{Experiment}
\label{sec:experiments}






In this section, we conduct experiments to demonstrate the robustness of SemaMark. In Section~\ref{exp:main_results}, we demonstrate that its robustness is better than the baseline methods. In Section~\ref{sec:quality}, we show that our watermark has almost no influence on the quality of generated texts. In Section~\ref{sec:alabtion}, we use ablation studies to demonstrate the effectiveness of partition function and $Q$-offset detection, and show the sensitivity of the partition function. In Section~\ref{exp:cl} we visualize the distribution of NE-Ring and provide analysis on the feature distribution of Contractive Learning.

\begin{table*}
\centering
\resizebox{0.97\textwidth}{!}{
\begin{tabular}{cl|cccc|cccc}
\hline
\multirow{2}{*}{} & \multicolumn{1}{c}{\multirow{2}{*}{Paraphrase}} & \multicolumn{4}{c}{ROC-AUC} & \multicolumn{4}{c}{F1 with best threshold} \\
& & LeftHash & SelfHash & EXP-Edit & ours & LeftHash & SelfHash & EXP-Edit & ours \\
\hline
\multirow{4}{*}{OPT-2.7B} & No paraphrase & 0.9913 & 0.9886 & 0.9799 & {0.9948} & {0.9921} & 0.9861 & 0.9708 & 0.9905 \\
& Translation & 0.9091 & 0.8147 & 0.8749 & \textbf{0.9692} & 0.8456 & 0.7622 & 0.8157 & \textbf{0.9330} \\
& Dipper & 0.9878 & 0.9728 & 0.9736 & \textbf{0.9911} & \textbf{0.9727} & 0.9400 & 0.9620 & {0.9701} \\
& GPT-3.5 & 0.9028 & 0.7908 & 0.9392 & \textbf{0.9406} & 0.8358 & 0.7378 & {0.8852} & \textbf{0.8902} \\
\hline
\multirow{4}{*}{OPT-6.7B} & No paraphrase & 0.9918 & 0.9930 & 0.9784 & {0.9949} & {0.9911} & 0.9863 & 0.9705 & 0.9858  \\
& Translation & 0.8807 & 0.8098 & 0.8625 & \textbf{0.9308} & 0.8129 & 0.7468 & 0.8013 & \textbf{0.8882}  \\
& Dipper & \textbf{0.9904} & 0.9747 & 0.9728 & {0.9871}  & {0.9786} & 0.9432 & 0.9620 & \textbf{0.9821} \\
& GPT-3.5 & {0.8990} & 0.7909 & 0.8996 & \textbf{0.9377} & 0.8300 & 0.7367 & 0.8354 & \textbf{0.8766} \\
\hline
\end{tabular}
}
\caption{\label{tab:main_results}
Watermark detection results under three paraphrases. (The best performance under paraphrase is bolded.)
}
\end{table*}

\subsection{Experiment setups}

\textbf{Backbone models and datasets.} We test our method by watermarking two models, OPT-2.7B and OPT-6.7B~\cite{zhang2022opt} which are referred to as the backbone models in following sections. For dataset, we use the news-like subset of C4~\cite{2020t5}, which covers a variety of topics. From the news-like subset of C4, we extract a training set, a validation set and a test set. For each sample, we use the first half of text as prompt to generate watermark sentences. More details can be found in Appendix~\ref{appd:baseline}.

\textbf{Baseline methods.}
We compare our method with three baselines  LeftHash, SelfHash~\cite{kirchenbauer2023reliability} and EXP-Edit~\cite{kuditipudi2023robust}. LeftHash and SelfHash are two methods based on the partition of vocabulary using the hashes of tokens. EXP-Edit uses a private sequence to encode the watermark by changing the probability distribution of the sequence of tokens. More details on the implementation can be found in Appendix~\ref{appd:baseline}.

\textbf{Paraphrase setups.}
We use three representative methods to paraphrase the watermarked text, round-trip translation~\cite{TiedemannThottingal:EAMT2020}, Dipper~\cite{krishna2023paraphrasing} and GPT-3.5.
For round-trip translation, we first translate from English to another language and then transform back to English, such that some words and expressions will be changed because the translation is not an one-to-one mapping. For Dipper, we follow the parameter setting in ~\citet{kirchenbauer2023reliability}. For GPT-3.5, we use the prompt in ~\citet{kirchenbauer2023reliability} to query GPT-3.5 for paraphrase. The examples of the three paraphrases can be found at Appendix~\ref{appd:parapgrase_ex}.

\textbf{Evaluation metrics and hyper-parameters.}
We use F1 score with best threshold and ROC-AUC to measure the performance of the watermark detection.
All the metrics are calculated based on at least 500 watermarked samples and 500 non-watermark samples. The length of watermarked samples before paraphrase and non-watermark samples is $200 \pm 25$.
In generation, we set $\gamma = 1/4$ for LeftHash, SelfHash and SemaMark. In detection, we set $\gamma = 1/3$ and $\delta = 2$ based on the validation set in Section~\ref{sec:alabtion}(b). In SemaMark, we set $m = 15$, $Q=15$, $K=5$ for OPT-2.7B and $K=4$ for OPT-6.7B.


\subsection{Main Results}
\label{exp:main_results}

In this subsection, we demonstrate the robustness of the proposed SemaMark under paraphrase by comparing it with three baseline methods on two backbone models. We first generate watermarked texts and use three paraphrase methods to remove the watermarks. The detection performance of both texts with and without paraphrase is reported in Table~\ref{tab:main_results}. As we can see, before paraphrase, all the watermarked methods have good detection performance. After paraphrase, SemaMark has the best detection performance most of the time across all the backbone models and all the paraphrase methods, which suggests that our method is more robust against paraphrase. 

In detail, by round-trip translation, the paraphrase reduces the detection ability of baseline methods effectively, while the watermark of SemaMark is robust. 
Under round-trip translation, the best ROC-AUC of baselines is 0.9091 on OPT-2.7B and 0.8807 on OPT-6.7B, respectively. But ROC-AUC of SemaMark is 0.9692 and 0.9308, which is at least 0.05 higher than all the baseline methods. Similarly, under paraphrase of GPT-3.5, SemaMark is better than all the baselines. The best baseline performance under GPT-3.5 is 0.9392 in ROC-AUC on OPT-2.7B and 0.8990 in ROC-AUC on opt-6.7B, but SemaMark has higher AUC-ROC of 0.9406 and 0.9377. For Dipper, we note that all methods are robust to Dipper since it does not significantly reduce the detection performance. However, SemaMark is still one of the most robust. On OPT-2.7B, it performs best in ROC-AUC, while on OPT-6.7B, it has the best F1 score. 
From Table~\ref{tab:main_results}, the results show an obvious improvement of SemaMark in robustness. This implies that using semantics as the seed for the partition function is effective under paraphrase.


\subsection{Text Quality}
\label{sec:quality}

\begin{figure}[ht!]
  \centering
  \begin{subfigure}[b]{0.235\textwidth}
    \includegraphics[width=\textwidth]{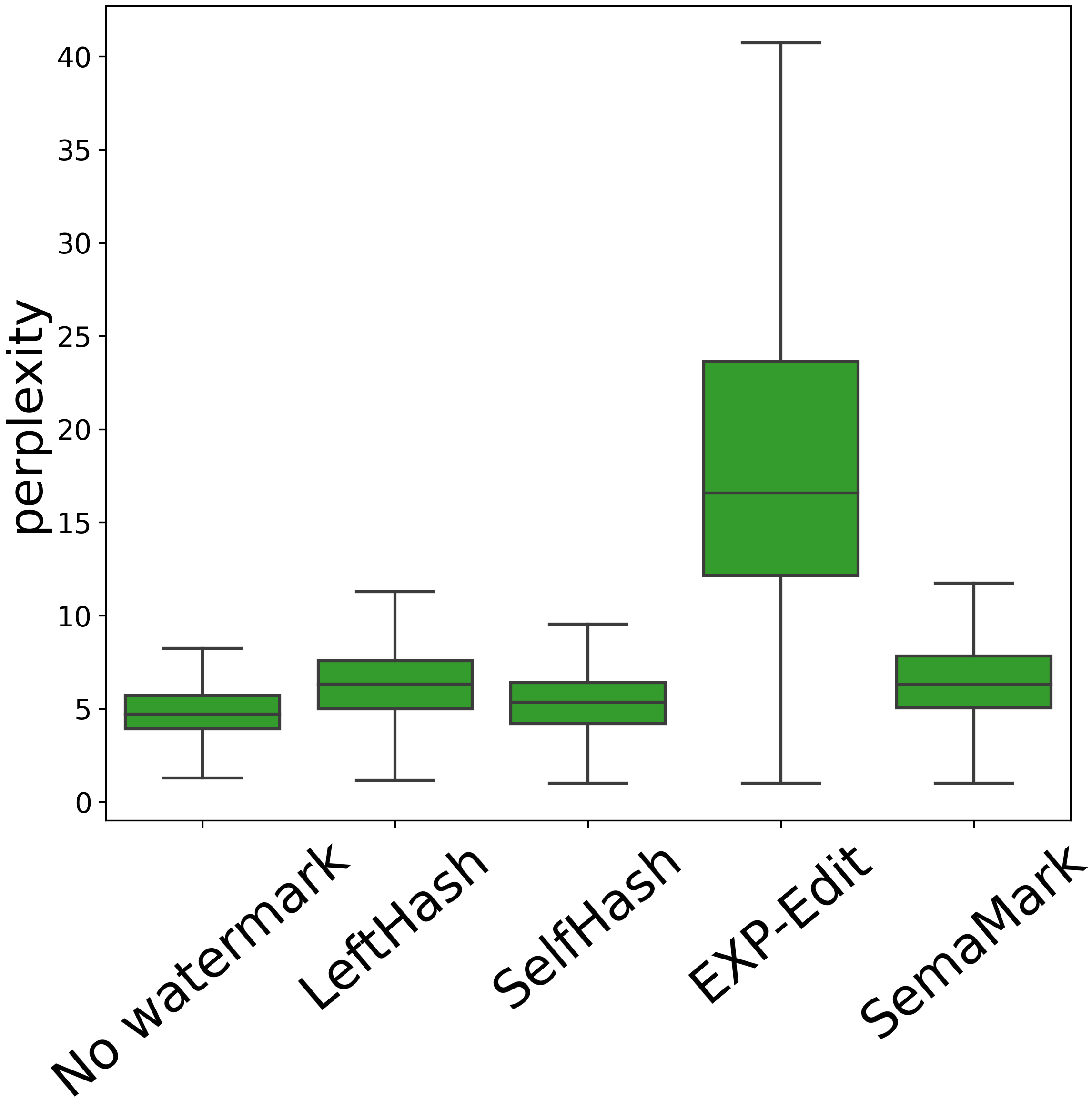}  
    \caption{OPT-2.7B}
    \label{fig:quality_sub1}
  \end{subfigure}
  \begin{subfigure}[b]{0.235\textwidth}
    \includegraphics[width=\textwidth]{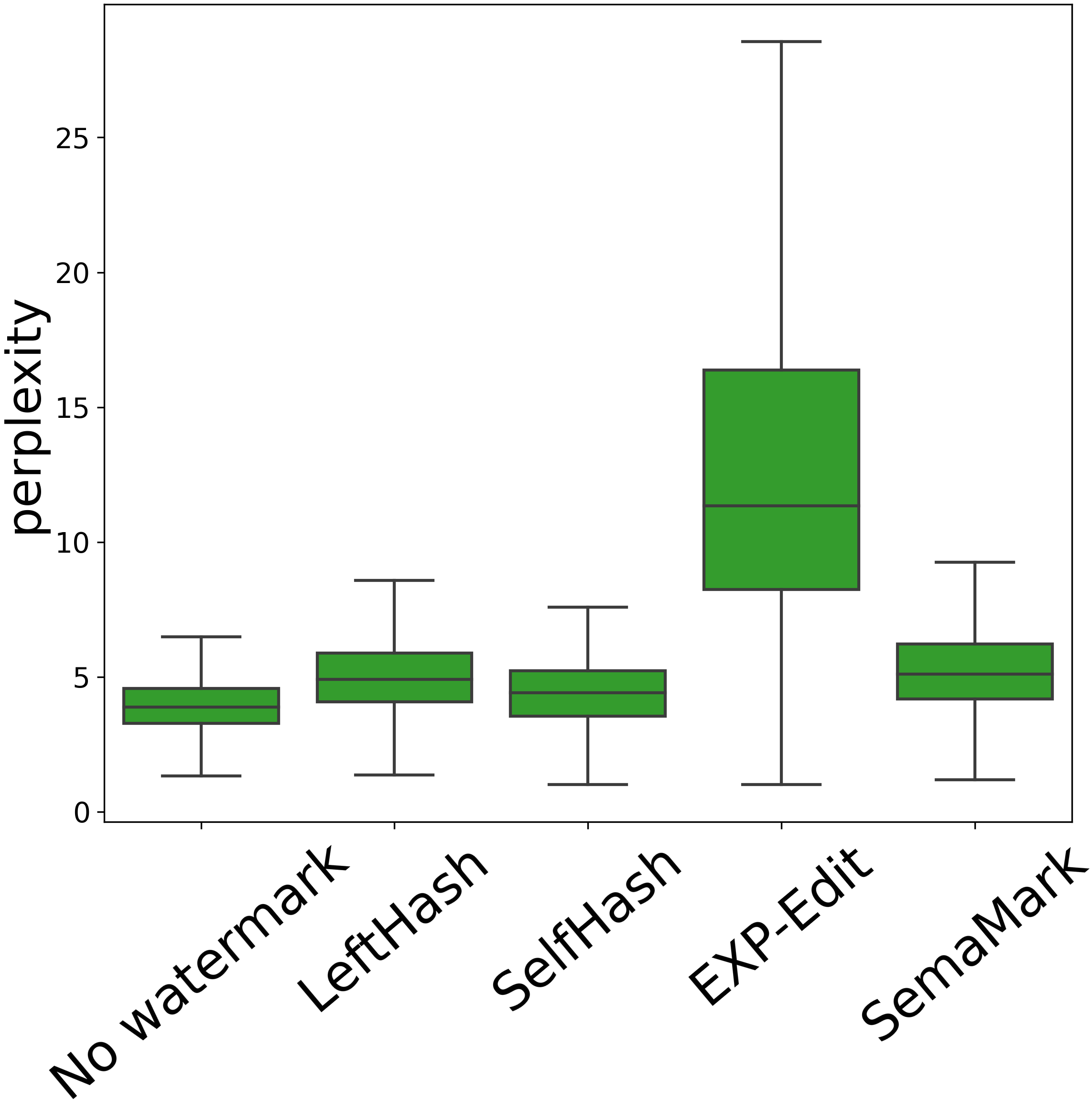}  
    \caption{OPT-6.7B}
    \label{fig:quality_sub2}
  \end{subfigure}
  \caption{Text quality (perplexity)}
  \label{fig:text_quality}
\end{figure}

\noindent Watermark should not compromise the generation quality of LLMs. In this subsection, we compare the text quality by calculating perplexity and demonstrate that our watermark has almost no influence on the generated quality. Perplexity measures the likelihood that a sentence is generated by one model. Lower perplexity means the watermarked text is more predictable. In other words, it is more consistent with the reasoning of the given model. In Figure~\ref{fig:text_quality}, we use OPT-6.7B with no watermark to get perplexity for all the watermarked methods. All the results in Figure~\ref{fig:text_quality} are calculated without paraphrase, because the generation quality of text is not related to paraphrase. From Figure~\ref{fig:quality_sub1} on OPT-2.7B, we can see that our watermark, LeftHash and SelfHash have almost no influence on the generation quality. They has perplexity at around 6 which is similar as the generated text without watermark. Instead, EXP-Edit has much higher perplexity, which means that EXP-Edit changes the generated text in an aggressive way and much reduces the generation quality after watermarking. This is probably because EXP-Edit adjusts the logits on the whole vocabulary. From Figure~\ref{fig:quality_sub2}, we can draw similar conclusions for OPT-6.7B. EXP-Exit also increases the perplexity by around 10, while the average perplexity of LeftHash, SelfHash and ours is around 1 higher than the non-watermarked generated text. In summary, our SemaMark can keep the quality and robustness simultaneously.


\subsection{Ablation Study}
\label{sec:alabtion}

In this subsection, we study the influence of the length of the sequence we use for generating one semantic value and the sensitivity of the partition function.

\begin{figure}[t]
  \centering
  \includegraphics[width=0.35
  \textwidth]{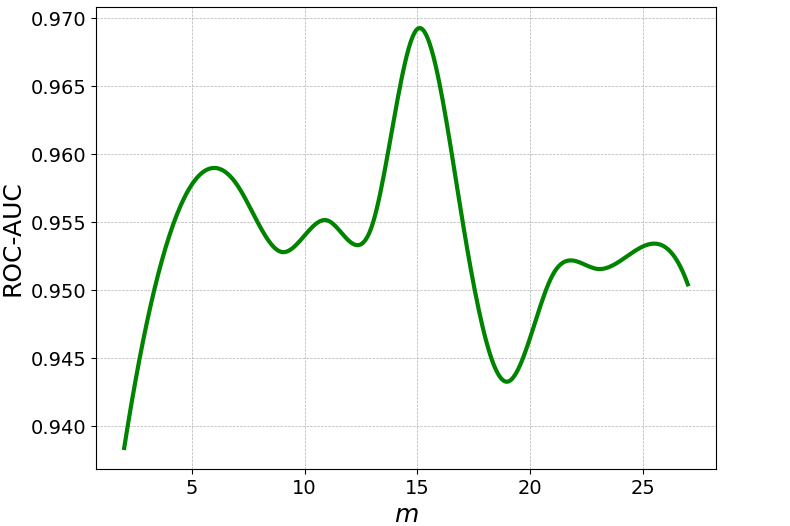}
  \caption{ROC-AUC and $m$}
  \label{fig:diff_m} 
\end{figure}

\paragraph{\textit{a)} Length of previous sequence tokens, $m$.} 

In the first step of SemaMark, i.e., \textit{weighted embedding pooling}, we use the semantic of the previous $m$ tokens to get the more stable embedding. But if the length of the sequence is too long, it will also hurt the robustness. In Figure~\ref{fig:diff_m}, we test watermark on OPT-2.7B with different $m$ and draw the ROC-AUC. The results show that before $m=15$, ROC-AUC is in the trend of increase as the $m$ changes. But when $m>15$, ROC-AUC becomes fluctuating. It is possibly because that the distant tokens will include more change after paraphrase as we mentioned in Section~\ref{sec:main_method}. Another possible reason is that in the beginning of generation for the first $m$ tokens, the number of previous tokens is smaller than $m$ and NE-Ring can only use the embeddings of limited tokens for prediction, which may be unstable. Thus, too long or too short sequence will hurt the robustness of SemaMark against paraphrase. In our experiments, we choose $m=15$ for all the settings.




\paragraph{\textit{b)} $Q$-offset detection}



\begin{figure}[t]
  \centering
  \begin{subfigure}[b]{0.235\textwidth}
    \includegraphics[width=\textwidth]{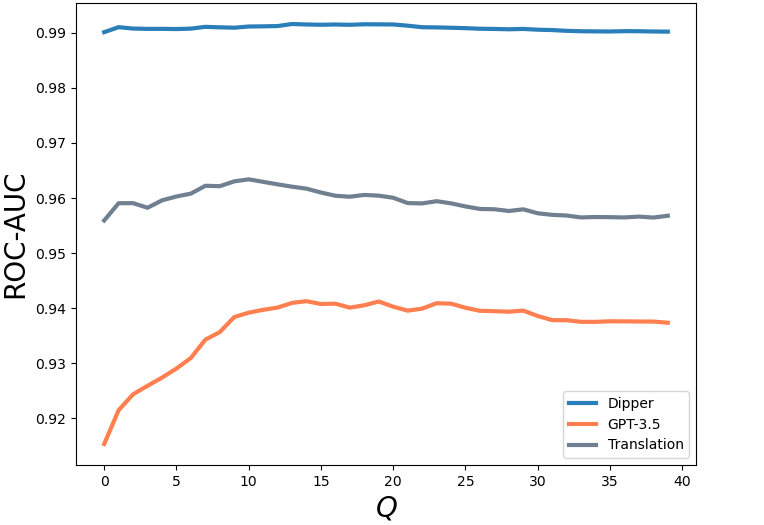}  
    \caption{ROC-AUC and offset $Q$}
    \label{fig:offset}
  \end{subfigure}
  \begin{subfigure}[b]{0.235\textwidth}
    \includegraphics[width=\textwidth]{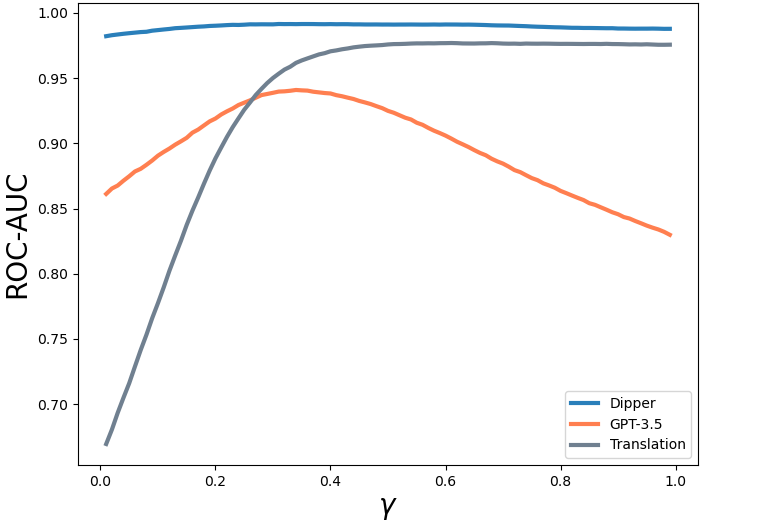}  
    \caption{ROC-AUC and $\gamma$}
    \label{fig:gamma}
  \end{subfigure}
  \caption{Text quality (perplexity)}
  \label{fig:q_offset_detection_abla}
\end{figure}

In this subsection, we show that the effectiveness of the proposed $Q$-offset detection. In Figure~\ref{fig:offset}, we demonstrate the change of ROC-AUC of SemaMark with different $Q$ in offset detection under three different paraphrases. $Q$-offset detection searches the highest $z$-statistics from $-Q$ to $Q$ as the $Q$-offset score. From Figure~\ref{fig:offset}, we can see that when $Q$ increases, ROC-AUC first increases and decreases after $Q$ is around 15. When $Q<15$, the offset can help correct the errors of semantic values close to the boundary. Compared with detection without offset, i.e. $Q=0$, ROC-AUC of SemaMark is much better, which means that the offset can help to solve the errors of semantic values around the boundaries that are more vulnerable to paraphrases.
When $Q>15$, the correction of this error is limited, because the offset will also increase the $Q$-offset score of negative samples as it also searches the highest $z$-statistics of negative samples. On the other hand, the computation cost will also increase if $Q$ is too large because it has to search more possible $q$. In practice, we set $Q=15$ in all the experiments, which can effectively reduce the influence of the errors of semantic values at the boundaries.

Since the $Q$-offset detection searches the highest green word fraction, the fraction of green list word of non-watermarked text will be higher than the $\gamma$ that we used to randomly select the green list. Thus, it is not accurate to use the original $\gamma$ for $z$-statistics. We treat $\gamma$ as a hyper-parameter and use a validation set to select its value. As shown in Figure~\ref{fig:gamma}, the detection performance of SemaMark under paraphrases of Dipper and GPT-3.5 will reach the highest when $\gamma$ is around $1/3$, while it will continue to increase under paraphrase. In practice, we set $\gamma = 1/3$ for $Q$-offset detection.

\paragraph{\textit{c)} Sensitivity of partition function.}

As we mentioned, the partition function is sensitive to any change of the input because it only uses the input as the seed of the random generator. To validate its sensitivity to continuous embeddings, we adopt the embedding vector as the input to show that, with tiny change of the embeddings, the partition of vocabulary can be very different. We propose a hash method based on \textsf{md5sum}~\cite{deepakumara2001fpga} to adopt the partition function by transforming the continuous embeddings to an integral seed. We use 1000 sequences to test the sensitivity. For each sequence embedding, we first get a green list from the partition function. Then we change one dimension of the embedding by only 1e-5 to get a new partition result. The overlapping of the green list before and after changing is 24.99\% on the average of 1000 sequences. It is consistent with $\gamma$ we use to watermark, because the random partition with the changed embedding is independent from the original one. It means the partition function is sensitive to any small change in its input. Instead, after we use NE-Ring to discretize the embeddings, the overlapping of green list after changing embeddings by 1e-5 is 100\%, which means the discretization can effectively handle this change. In practice, SemaMark can provide the tolerance that is much larger than 1e-5, which makes the watermark more robust under paraphrase. With the improvement of $Q$-offset, the detection of SemaMark is more robust and effective.

\paragraph{\textit{d)} Model size.} To show the robustness of our method on different model sizes, in this section, we also test the watermark under round-trip translation paraphrase on LLaMA-7B and LLaMA2-7B, which have larger size and different architectures. As indicated in Table 3, our approach consistently exhibits the highest robustness against paraphrasing. Specifically, in the LLaMA2-7B model, SemaMark significantly outperforms the baseline models, achieving an increase of 0.06 and 0.03 in ROC-AUC. Similarly, in the LLaMA-7B model, our method shows superior performance with an increase of 0.027 and 0.009 in ROC-AUC.
\begin{table}
\centering
\resizebox{0.48\textwidth}{!}{
\begin{tabular}{c|ccc|ccc}
\hline
 & \multicolumn{3}{c}{ROC-AUC} & \multicolumn{3}{c}{F1 with best threshold} \\
& LeftHash & SelfHash & ours & LeftHash & SelfHash & ours \\
\hline
LLaMA-7B & 0.819 & 0.838 & 0.846 & {0.748} & {0.774} & 0.781 \\

LLaMA2-7B & 0.811 & 0.841 & 0.872 & 0.749 & 0.773 & 0.810 \\

\hline
\end{tabular}
}
\caption{\label{tab:model_size}
Watermark detection results under different model size.)
}
\end{table}


\subsection{Distribution on NE-Ring based on CL}
\label{exp:cl}

\begin{figure}[ht!]
  \centering
  \begin{subfigure}[b]{0.23\textwidth}
    \includegraphics[width=\textwidth]{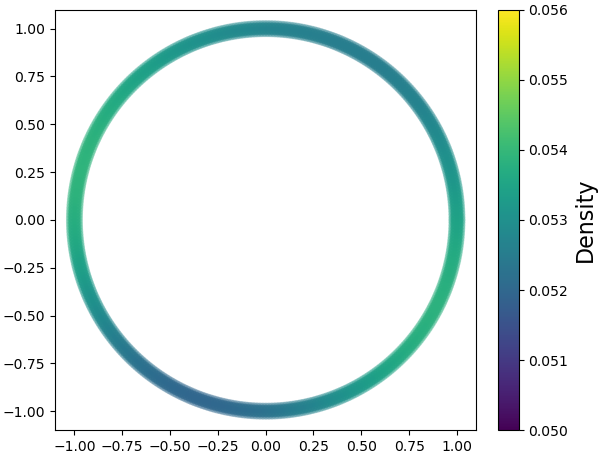}  
    \caption{NE-Ring}
    \label{fig:sub1}
  \end{subfigure}
  \begin{subfigure}[b]{0.23\textwidth}
    \includegraphics[width=\textwidth]{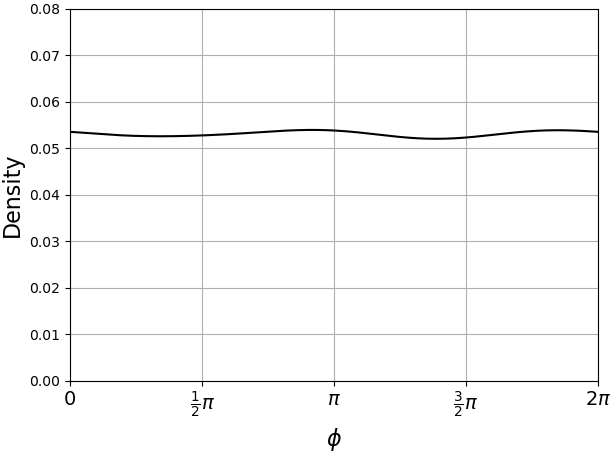}  
    \caption{Distribution on $\phi$}
    \label{fig:sub2}
  \end{subfigure}
  \caption{Visualization of NE-Ring}
  \label{fig:test}
\end{figure}

In this subsection, we demonstrate that Contrastive Learning can help evenly distribute the semantics on the NE-Ring. The even distribution can help the sequences reach all  possible semantic values and provide more diverse semantic values to prevent the watermark from being cracked by counting token frequency. In Figure~\ref{fig:sub1}, we use Gaussian density estimation~\cite{chen2017tutorial} to get the distribution of the semantics on the NE-Ring \textbf{before} discretization. We use different colors to show the density. The NE-Ring in Figure~\ref{fig:sub1} shows that, the distribution is uniform. All the density is between 0.052 and 0.054. We further plot the density based on the polar angle $\phi$ in Figure~\ref{fig:sub2} where the density has almost no change on all the polar angle from 0 to $2\pi$. This implies that the training based on Contrastive Learning can ensure the semantics will reach all possible discrete values. It can prevent the case where the discrete values will gather in some discrete sections and produce monotonous vocabulary partitions. As a result, it can protect the watermark from being cracked by counting token frequency.

\section{Conclusion}

In this paper, we use the semantic information for watermarking to enhance the robustness against paraphrase. The existing watermark methods use the matching between the previous tokens and the partition vocabulary. This matching can be easily broken by paraphrase. However, we construct the mapping between the semantics and the vocabulary. In this way, the semantics will stay stable under paraphrase and the robustness of watermark can be enhanced. To make use of semantics, we propose SemaMark to discretize the embedding space on NE-Ring and propose a training method based on CL. In addition, we use $Q$-offset detection to further advance the robustness by increasing the tolerance of the semantic values close to the discrete boundary. In experiments, we demonstrate our method can perform much better compared with baseline methods under paraphrase while having little influence on the generation quality.

\section{Limitations}

In some cases, the customers may rely on some API-based LLMs and do not have the access to the embeddings and the permission to modify the logits during generation. Although our watermark method can effectively detect the LLM-generated content and increase the detection success rate under paraphrase, it is not applicable for black-box LLMs. The second weakness of our method is that the NE-Ring is dependent on the semantic embedding of LLMs. For each LLM, we need to train a specialized EN-Ring, which might be inflexible if we want to produce a general model for NE-Ring or fine-tune the LLMs. Despite the weaknesses, our method is successful in the problem of robustness under paraphrase. In the future work, we will continue to extent our method into black-box LLMs and a universal model that does not require customized training for various specific LLMs.

\textbf{Potential risk.} Our discussion about the robustness might provide motivation for the attackers to find other methods like adaptive attack. Although we provide robustness under paraphrase, if the unauthorized people propose possible attack method focusing on the green-list based watermark from other perspectives, the detection rate for LLM-generated texts are still possible to be reduced. 

\section{Acknowledgements}

Jie Ren, Han Xu, Yingqian Cui, and Jiliang Tang are supported by the National Science Foundation (NSF) under grant numbers CNS 2246050, IIS1845081, IIS2212032, IIS2212144, IOS2107215, DUE 2234015, DRL 2025244 and IOS2035472, the Army Research Office (ARO) under grant number W911NF-21-1-0198, the Home Depot, Cisco Systems Inc, Amazon Faculty Award, Johnson\&Johnson, JP Morgan Faculty Award and SNAP.

\bibliography{anthology,custom}

\newpage

\appendix

\begin{table*}
\centering
\resizebox{0.95\textwidth}{!}{
\begin{tabular}{c|p{6.5cm}|p{6.5cm}}
\hline
 & Original & Paraphrased \\
\hline
Translation & The course ran from Feb. 16 to Feb. 18 and was designed to address officers’ response to critical incidents and to enhance officer safety. “I am excited by the training and professionalism demonstrated by the officers at the training session. They were receptive and focused on learning new techniques and approaches to policing.” Commissioner Walwyn added. & The training, which was held from 16 February to 18 February, was aimed at discussing police responses to major incidents and enhancing police security. " I am excited about the training and professionalism shown by police officers in the course, who are willing to accept and focus on new policing techniques and methods. " Commissioner Walwyn adds. \\
\hline

GPT & She tells him she wants to protect him. She asks what else she can do. Stefan tells her not to trust Chad. She vows she won’t. Stefan offers to get rid of Chad for her if she wants. She says she wants Chad and his family gone. She wants to have his babies. She tells him to get out of her way. She doesn’t know if she can love him, but she wants his babies. She wants to be a mother. Stefan tells her she deserves a real man. He’ll love her and protect her. She says she wants him and his babies. She wants to have his babies. Stefan says she deserves a real man. He’ll love her and protect her. She says she wants him and his babies. She wants to have his babies. Stefan says she deserves a real man. He’ll love her and protect her. She says she wants him and his babies. & According to Reuters, China has already taken steps to solidify its presence in the South China Sea. It has constructed an airstrip on one island and In a tense conversation, a woman expresses her desire to protect a man, Stefan, and seeks his advice on how to do so. Stefan advises her not to trust a person named Chad, to which she firmly pledges her distrust. Stefan offers to eliminate Chad from her life if she desires, and she responds by expressing her wish for Chad and his family to disappear entirely. In a surprising revelation, she declares her desire to have Stefan's children, emphasizing her longing to become a mother. Despite uncertain feelings of love towards Stefan, she insists on wanting his babies and urges him to step aside. Stefan, in turn, assures her that she deserves a genuine partner who will love and safeguard her. Yet, undeterred, she asserts her desire for both Stefan and his offspring, reiterating her ambition to have his babies. \\
\hline

DIPPER & Kewanee got the ball again with 7.2 seconds to go, but couldn’t get a shot off. “I think it was something we needed to have a plan for and it was one that worked in the end. We did it well enough,” Keene said. “We were trying to get to the rim. A lot of times when you are in a situation like that and you can’t score, it’s better to just foul.”Macomb was 12-for-27 from behind the arc (42 percent). The Grinnell offense is centered around perimeter shooters. The two perimeter players for Macomb are Emerick and Reilly Rieder. The next closest shooter on the team is Rieder with 10 points. Nolan led Kewanee with 15 points. Macomb, now 14-5, now travels to Jacksonville on Thursday to play Jacksonville & Kewanee had the ball again with 7.2 seconds left but couldn’t get off a shot. “I think that’s something we needed to have a plan for and in the end, it worked,” coach Lambert said. “We did it well enough,” Keene said. “We were trying to get to the rim. Often when you are in a situation like that and you can’t score, it’s better to foul.” Macomb was 12 for 27 from beyond the arc (42 percent). The Grinnell offense is based on sharpshooting players. Macomb's two shooters are Emerick and Rieder. Rieder has ten points. Nolan led Kewanee with 15 points. Macomb, now 14-5, will play at Jacksonville Thursday. \\
\hline
\end{tabular}
}
\caption{\label{tab:paraphrase_ex}
Paraphrase examples.
}
\end{table*}

\section{More details on experimental settings}
\label{appd:baseline}

All the baseline models, backbone models and datasets we use are open source and available for academic purpose.
For backbone models, we use the open-sourced model from Huggingface\footnote{https://huggingface.co/facebook/opt-2.7b}. The implementation is based on Pytorch\footnote{https://pytorch.org/} framework and also depend on packages including NLTK~\cite{bird2009natural} and Numpy~\cite{harris2020array}. For baseline methods, we use the released official code from the authors. For paraphrase models, we use OPUS-MT translation model and Dipper on Huggingface repository\footnote{https://huggingface.co/Helsinki-NLP/opus-mt-en-zh and https://huggingface.co/kalpeshk2011/dipper-paraphraser-xxl}, and API of ChatGPT\footnote{https://chat.openai.com/chat}.

\section{Examples of paraphrases}
\label{appd:parapgrase_ex}

In Table~\ref{tab:paraphrase_ex}, we provide the examples of three paraphrases.


\end{document}